\documentstyle[twoside,wicsbook,graphicx]{article}

\begin{document}

\title{\vspace{-3pc}%
\titlesize{\bf Neural Networks for Spectral Analysis of
Unevenly Sampled Data}}
\author{{\large Roberto Tagliaferri and Angelo Ciaramella} \\
{\normalsize Dipartimento di Matematica ed Informatica, }\\
{\normalsize Universit\`{a} di Salerno, and INFM unit\`{a} di Salerno,}\\
{\normalsize via S. Allende, 84081 Baronissi (SA) Italia }\\
{\normalsize and IIASS ''E. R. Caianiello'', Vietri s/m Italia} \and 
{\large Leopoldo Milano and Fabrizio Barone} \\
{\normalsize Dipartimento di Scienze Fisiche, Universit\`{a} di Napoli
''Federico II'' }\\
{\normalsize Istituto Nazionale di Fisica Nucleare, sez. Napoli, }\\
{\normalsize Complesso Universitario di Monte Sant'Angelo, }\\
{\normalsize via Cintia, I-80126 Napoli Italia}\\
}
\maketitle

\thispagestyle{empty}

\begin{abstract}
\ninesize
\noindent In this paper we present a neural network based estimator system
which performs well the frequency extraction from unevenly sampled signals. It
uses an unsupervised Hebbian nonlinear neural algorithm to extract the
principal components which, in turn, are used by the MUSIC frequency
estimator algorithm to extract the frequencies.

We generalize this method to avoid an interpolation preprocessing step and
to improve the performance by using a new stop criterion to avoid
overfitting.

The experimental results are obtained comparing our methodology with the
others known in literature.
\end{abstract}



\section{Introduction}

Periodicity analysis of unevenly collected data is a relevant issue in
several scientific fields. Classical spectral analysis methods are
unsatisfactory to solve the problem. In this paper we present a neural
network based estimator system which performs well the frequency extraction
from unevenly sampled signals. It uses an unsupervised Hebbian nonlinear
neural algorithm to extract the principal components of the signal
auto-correlation matrix, which, in turn, are used by the MUSIC frequency
estimator algorithm to extract the frequencies \cite{Khar94}, \cite{rasile et al}, \cite{Tagliaferri 99}. We generalize this method to
avoid an interpolation preprocessing step, which generally adds high noise
to the signal, and improve the system performance by using a new stop
criterion to avoid overfitting problems. The experimental results are
obtained comparing our methodology with the others known in literature (see \cite{Kay}, \cite{Marple}, \cite{Opp}, \cite{Lomb76}, \cite{Scargle}, 
\cite{Ferraz-Mello}).

\section{Evenly and unevenly sampled data}

In what follows, we assume $x$ to be a physical variable measured at
discrete times $t_{i}$. ${x(t_{i})}$ can be written as the sum of the signal 
$x_{s}$ and random errors $R$: $x_{i}=x(t_{i})=x_{s}(t_{i})+R(t_{i})$. The
problem we are dealing with is how to estimate fundamental frequencies which
may be present in the signal $x_{s}(t_{i})$ 
\cite{Deeming}, \cite{Kay}, \cite{Marple}.

If $X$ is measured at uniform time steps (even sampling) 
there are a lot of tools to effectively solve
the problem which are based on Fourier analysis 
\cite{Kay}, \cite{Marple}, \cite{Opp}. These methods, however, are usually
unreliable for unevenly sampled data \cite{Horne}. For instance, the typical approach of
resampling the data into an evenly sampled sequence, through interpolation,
introduces a strong amplification of the noise which affects the
effectiveness of all Fourier based techniques which are strongly dependent
on the noise level. 

To solve the problem of unevenly sampled data, we consider two classes of
spectral estimators:

Spectral estimators based on Fourier Trasform (Least Squares methods);

Spectral estimators based on the eingevalues and eingevectors of the
covariance matrix (Maximum Likelihood methods).

Classic Periodogram \cite{Kay}, \cite{Marple}, \cite{Opp}, Lomb's
Periodogram \cite{Lomb76}, Scargle's Periodogram \cite{Scargle}, DCDFT  
\cite{Ferraz-Mello} are the methods of the first class that we use, while
MUSIC \cite{Kay}, \cite{Marple}, and ESPRIT \cite{Roy} belong to second class.

The methods based on the covariance matrix are more recent and have great
potentiality. Starting by this consideration, we develop a method based on
the MUSIC estimator. It is compared with classic methods to highlight the
results.

\section{The Neural Estimator}

In the last years several papers dealed with learning in PCA neural nets 
\cite{Oja82}, \cite{sanger}, \cite{Khar94}, \cite{rasile et al}, \cite{Tagliaferri 99} finding advantages, problems and
difficulties of such neural networks. In what follows we shall use a robust
hierarchical learning algorithm 
\begin{eqnarray}
{\bf w}_{k+1}(i) &=&{\bf w}_{k}(i)+\mu _{k}g(y_{k}(i)){\bf e}_{k}(i), \\
\qquad {\bf e}_{k}(i) &=&{\bf x}_{k}-\sum_{j=1}^{i}y_{k}(j){\bf w}_{k}(j)
\label{eq34}
\end{eqnarray}

where ${\bf w}_{k}(i)$ is the weight vector of the $i-th$ output neuron at
step $k$, $y_{k}(i)$ is the corresponding output, $\mu _{k}$ is the learning
rate and $g(t) =\tanh \left( \alpha t\right) $ is learning
function because it has been experimentally shown that it is the best performing one in our problem \cite{rasile et al}, \cite{Tagliaferri 99}.

Our neural estimator (ne) can be summarized as follows:

\begin{itemize}
\item[1]  Preprocessing: calculate and subtract the average pattern to
obtain zero mean process with unity variance.

Interpolate input data if it is the case.

\item[2]  Initialize the weight matrix and the other neural network
parameters;

\item[3]  Input the $k-th$ pattern ${\bf x}_{k}=[x(k),\ldots ,x(k+N+1)]$ where 
$N$ is the number of input components.

\item[4]  Calculate the output for each neuron $y(j)={\bf w}^{T}(j){\bf x}%
_{i}\qquad \forall i=1,\ldots ,p$.

\item[5]  Modify the weights ${\bf w}_{k+1}(i)={\bf w}_{k}(i)+\mu
_{k}g(y_{k}(i)){\bf e}_{k}(i)\qquad \forall i=1,\ldots ,p$.

\item[6]  If convergence test is true then goto {\bf STEP 8}.

\item[7]  $k=k+1$. Goto {\bf STEP 3}.

\item[8]  End.

\item[9]  Frequency estimator: we use the frequency estimator MUSIC. It
takes as input the weight matrix columns after the learning. The estimated
signal frequencies are obtained as the peak locations of the function of
following equation \cite{Kay}, \cite{Marple}, \cite{Tagliaferri 99}: 
\[
P_{MUSIC}=\frac{1}{M-\sum_{i=1}^{M}|{\bf e}_{f}^{H}{\bf w}(i)|^{2}} 
\]
\end{itemize}

where ${\bf w}(i)$ is the $i-$th neural network weight vector after learning, and ${\bf e}_{f}^{H}$ is the pure sinusoidal vector. 
In the case of an interpolation preprocessing ${\bf e}_{f}^{H}=[1,e_{f}^{j2\pi f},\ldots ,e_{f}^{j2\pi f(L-1)}]^{H}$. In the generalization to non interpolated input data, ${\bf e}_{f}^{H}=[1,e_{f}^{j2\pi ft_{0}},\ldots
,e_{f}^{j2\pi ft_{(L-1)}}]^{H}$ where 
$\left\{ t_{0},t_{1},...,t_{\left(L-1\right) }\right\}$ are the first $L$ components of the temporal coordinates of the uneven signal.

When $f$ is the frequency of the $i-$th sinusoidal component, $f=f_{i}$, we
have ${\bf e}={\bf e}_{i}$ and $P_{MUSIC}\rightarrow \infty $. In practice
we have a peak near and in corrispondence of the component frequency.
Estimates are related to the highest peaks

Furthermore, to optimize the performance of the PCA neural networks, we stop
the learning process when $\sum_{i=1}^{p}|{\bf e}_{f}^{H}{\bf w}%
(i)|^{2}<M$ $\forall f$, so avoiding overfitting problems. In fact
leaving the stop condition used in the ne causes to the ne to find
periodicities not present in the signal, while the new condition preserves
it from this problem. A simple example is illustrated in figure 2 where we
can see how the frequency identification varies depending on the stop
condition (see next section for signal information). In fact, without the early stopping (see figure 2.b), $\Delta f$ after $100$ epochs remains at a value between $0.20$ and $0.25$ and we cannot know when the system reaches the best performance. The new stopping criterion, instead, permits to $\Delta f$ to have a final value about $0.0$ just after $50$ epochs (see figure 2.a).

\section{Experimental Results}

Many experiments on synthetic and real signals were made, and in this paper
we present the results obtained with one specific real signal, which
highlights the main features of our problem.

The real signal is related to the Cepheid SU Cygni \cite{Fernie}.
The sequence was obtained with the photometric technique UBVRI and the
sampling made from June to December 1977. The light curve is composed by 21
samples, and has a period of $3.8^{d}$, as shown in figure 1.

The first experiment is concerning the interpolation. In this case we apply
three different methods by using the Signal Processing Mathlab $^@$ Toolbox: linear, cubic and spline, because they are quite simple and the most
used ones. In figure 3 there is a plotting of the interpolating functions and the frequency estimates obtained by the ne with the spline interpolated signal as input.

In this case, the parameters of the ne are: $N=10$, $p=2$, $\alpha =20$, 
$\mu =0.001$. The estimate frequency interval is $\left[ 0(1/JD),0.5(1/JD)%
\right] $. The estimated frequency without interpolation is $0.260$ (1/JD).

A comparison is made with the other methods cited in a previous section and
the experimental results are shown in figure 4 and in table 1. Only the
Lomb's Periodogram is in agreement with the right periodicity, but showing
some spurious peaks. Furthermore, if we enlarge the frequency window for the
two best performing methods, while the ne continues to work well, the Lomb's
periodogram does not work at all as illustrated in figure 5.

\section{Concluding Remarks}

In this paper we have illustrated an improved technique based on PCA neural 
Networks and MUSIC to estimate the frequency of unevenly sampled data. 
It has been shown that it obtains good results on real data (here we used the SU 
Cygni light curve) compared with other well-known methods.
In fact, it obtains a good estimate of the signal frequency also with few unevenly sampled inputs, it reduces the noise problems related to input data interpolation, 
it optimizes the convergence by introducing an early stopping criterion, and, finally, it is more resistant to the dimension of the frequency windows.

Future research lines regard the introduction of genetic algorithms to optimize 
the weight initialization of
the PCA neural networks and to use filters to extract and identify one frequency 
at each time when dealing with multi-frequency signals.

\begin{figure}
\begin{center}
\includegraphics{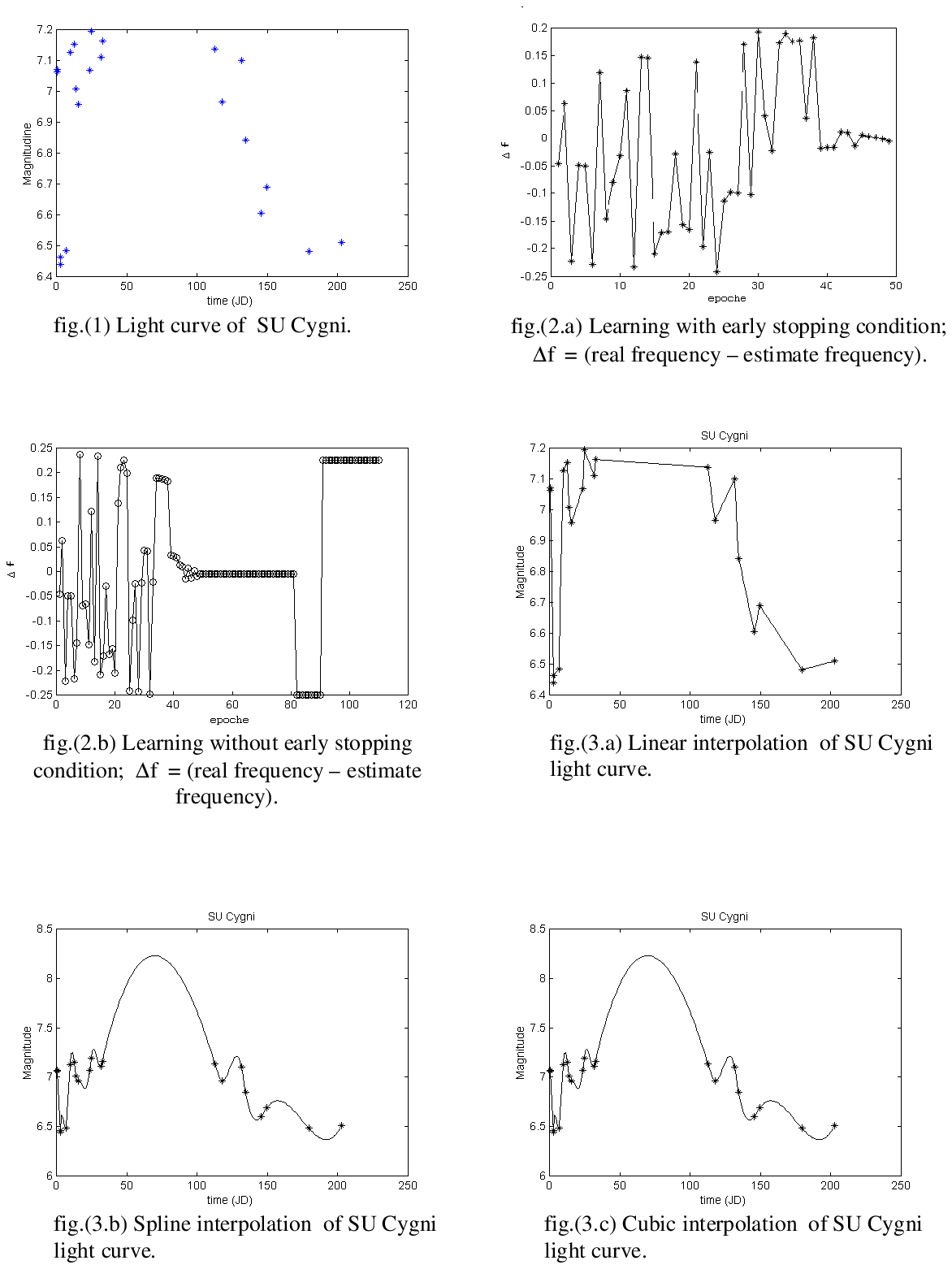}
\end{center}
\end{figure}

\begin{figure}
\begin{center}
\includegraphics{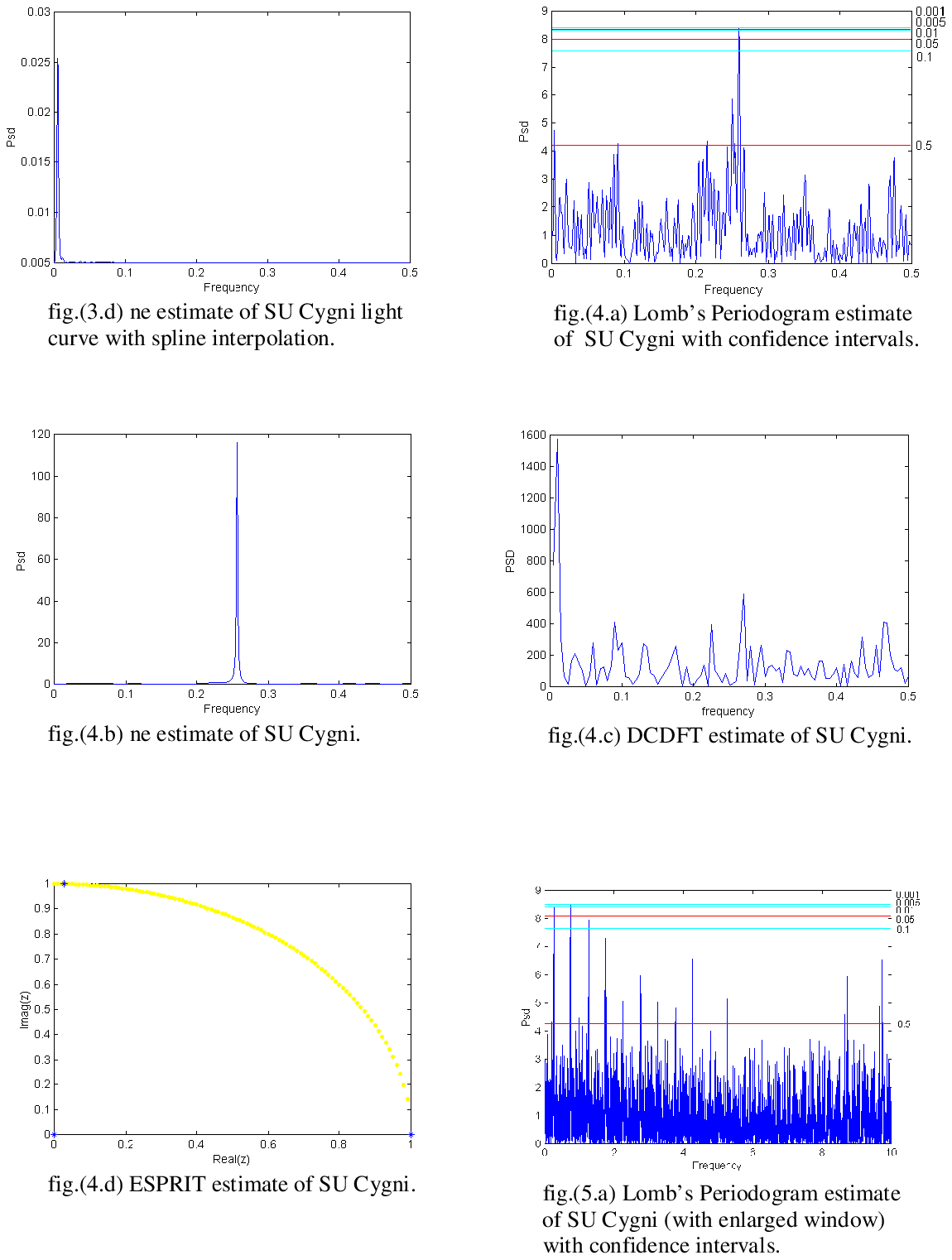}
\end{center}
\end{figure}

\begin{figure}
\begin{center}
\includegraphics{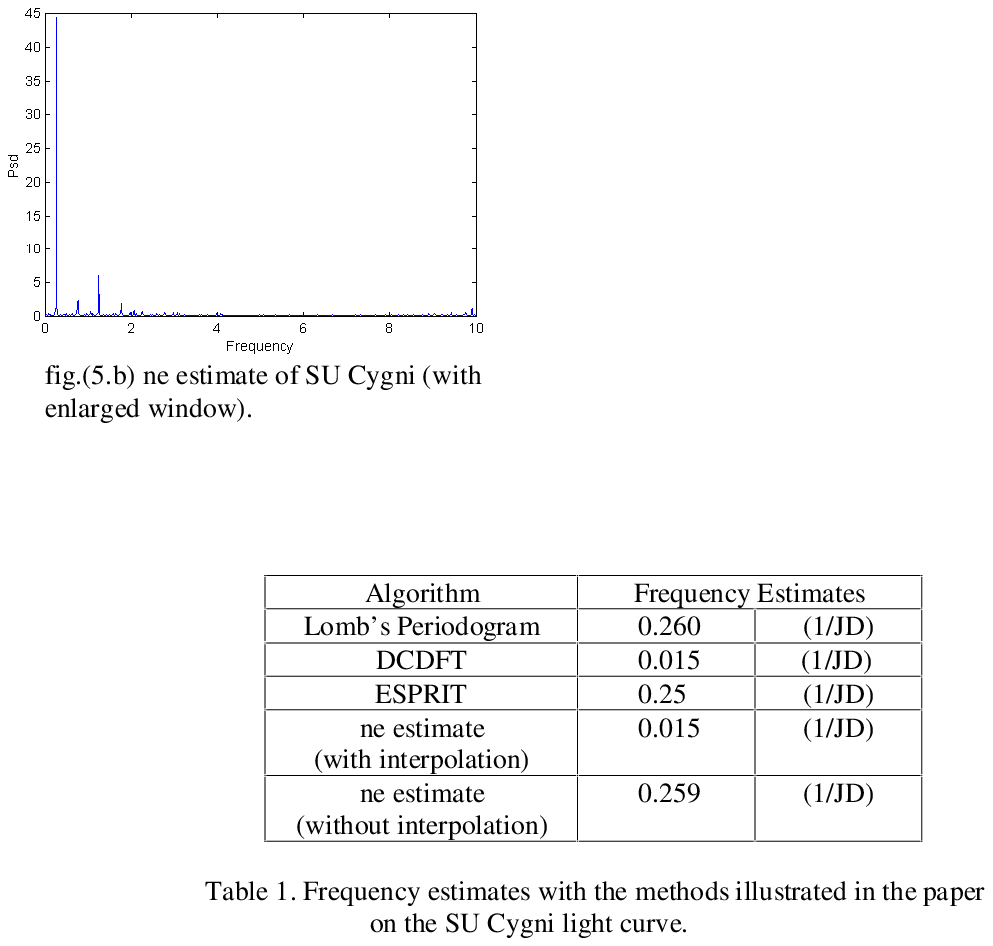}
\end{center}
\end{figure}

\end{document}